\documentclass[11pt,twoside]{article}

\usepackage{asp2006}
\usepackage{epsf}
\usepackage{psfig}
\usepackage{lscape}

\begin{document}
\title{Mapping of the cosmic ray events related to the solar activity for the period 2003-2005}   
\author{A. Papaioannou, P. Makrantoni, H. Mavromichalaki}   
\affil{Nuclear and Particle Physics Section, National and Kapodistrian University of Athens, Pan/polis 15771 Athens Greece }    

\begin{abstract}
The relationship between cosmic ray intensity decreases and solar events is still an open field of space research. 
In this work a complete study of solar events occurred from January 2003 to December 2005, is considered. 
This three-years time period characterized by an unexpected activity of the Sun was divided into 27-day intervals starting from 
BR 2313 (06.01.2003) to 2353 (21.12.2005), generating diagrams of the cosmic ray intensity data recorded at the Athens Neutron Monitor 
Station. This station is working at an altitude of 260m and cut-off rigidity 8.53GV provided to the Internet high-resolution data in real-time. 
A mapping of all available solar and interplanetary events, such as solar flares with importance M and X, coronal mass ejections (Halo and Partial) 
was done. As we are going down from the solar maximum to the declining phase of the 23rd solar cycle, a statistical overview of the corresponding 
relationship among these phenomena, the significant percentage of the connection of Halo CMEs and solar flares and the respective connection to Forbush 
decreases on yearly and monthly basis are discussed. The close association, as well as a probable quantitative analysis, between solar events is being 
denoted. The role of extreme solar events occurred in October / November 2003 and January 2005 is also discussed. Obtained results may be useful for 
predictions of transient solar events and space weather forecasting.
\end{abstract}

\section{Introduction}
As the Sun is the driver of space weather, solar events such as solar flares, coronal mass ejections etc. are close related to the Forbush decreases (FDs) 
recorded at the ground based neutron monitor stations \citep{Cane2000}. Recent studies verified this aspect \citep{Mavromichalaki2005}. 

\section{Data selection}
Cosmic ray measurements in Athens initiated in November 2000 with a standard super 6NM-64 neutron monitor \citep{Mavromichalaki2009}. 
Athens NM data are available in the on-line database at: http://cosray.phys.uoa.gr.
CME Lists of the U.S. Naval Research Laboratory (NRL) on the Large Angle and Spectrometric Coronagraph (LASCO) are used. 
These lists represent a subset of the final LASCO dataset and can be accessed through the web at the site: http://cdaw.gsfc.nasa.gov/. 
Data for solar flares were taken from: 
http://www.\\ngdc.noaa.gov/stp/SOLAR/ftpsolarflares.html. 

\section{Mapping of  Solar Activity on CR data}
\begin{figure}[!ht]
\begin{center}
\includegraphics{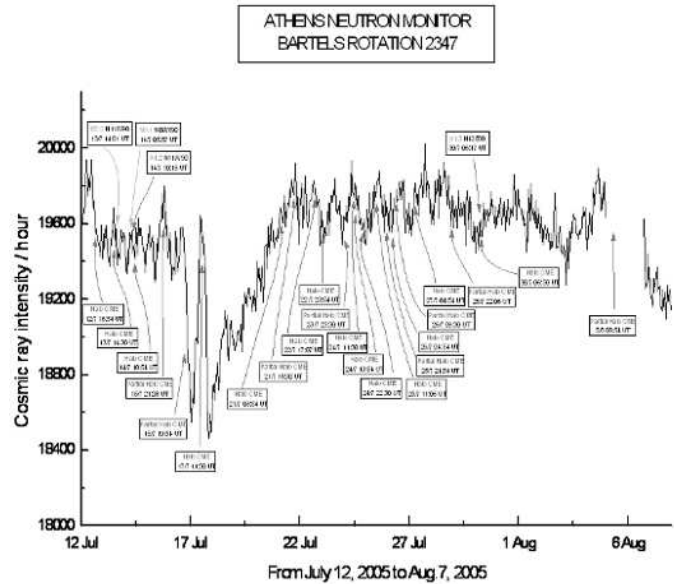}
\end{center}
\caption{Mapping of solar activity on cosmic ray measurements for BR 2347}\label{Papaioannou_A_fig1}
\end{figure}

 On the constructed  diagrams all available data, such as: time of first observation, date the event occurred and its co-ordinates are recorded (Fig. \ref{Papaioannou_A_fig1}). 
 In this way  an overall statistical picture of the relation between CMEs, SFs and FD is presented. 

 \section{Discussion}
 All major events as those of October-November 2003, January 2005, July 2005 \citep{Papaioannou2009a} and August-September 2005 \citep{Papaioannou2009b} 
 were succesfully recorded. In most of these cases cosmic rays revealed crusial information on the development of the events.

\end{document}